\documentclass[final,1p,times]{elsarticle}

\usepackage{amssymb}
\usepackage{amsthm}

\usepackage{threeparttable} 
\usepackage{caption}
\usepackage{subcaption}

\usepackage{pmboxdraw}
\usepackage{fancyvrb}

\usepackage{dirtree}
\usepackage{amsmath}
\usepackage{mathabx}
\usepackage[hidelinks]{hyperref}
\usepackage{multirow}
\usepackage{siunitx}
\usepackage{adjustbox}
\usepackage{listings}
\usepackage{color}
\usepackage{caption}

\definecolor{bg}{rgb}{0.95,0.95,0.95}
\definecolor{mygray}{rgb}{0.5,0.5,0.5}
\definecolor{mygreen}{rgb}{0, 0.6, 0.0}
\definecolor{mymauve}{rgb}{0.58,0,0.82}
\definecolor{background}{RGB}{39, 40, 34}
\definecolor{string}{RGB}{230, 219, 116}
\definecolor{comment}{RGB}{117, 113, 94}
\definecolor{normal}{RGB}{248, 248, 242}
\definecolor{identifier}{RGB}{166, 226, 46}

\newcommand{\specialsize}{\fontsize{7.6}{9}}

\newcommand{\etal}{\textit{et~al}. }

\newcommand{\apost}[1]{`#1'}
\newcommand{\postem}[1]{\emph{\apost{#1}}}
\newcommand{\bifoldpy}{\postem{bifold.py}}
\newcommand{\simpsonpy}{\postem{simpson.py}}
\newcommand{\filonpy}{\postem{filon.py}}

\DeclareUnicodeCharacter{0301}{\'{e}}
\DeclareUnicodeCharacter{03B1}{$\alpha$}

\newcounter{bla}

\journal{Computer Physics Communications}

\begin{document}

\begin{frontmatter}






\title{BiFold: A Python code for the calculation of double folded (bifold) potentials with density-in/dependent nucleon-nucleon interactions}



\author{\href{https://scholar.google.com/citations?user=dORZ704AAAAJ}{Mesut Karakoç}\corref{author}}

\cortext[author] {Corresponding author.\\\textit{E-mail address:} karakoc@akdeniz.edu.tr or mesutkarakoc@gmail.com}
\address{Department of Physics, Faculty of Science, \\Akdeniz University, TR 07070, Antalya, Turkey}

\begin{abstract}
BiFold calculates the density-dependent (DDM3Y$n$, BDM3Y$n$, CDM3Y$n$) or independent double folded potentials between 
two colliding spherical nuclei.
It is written in a Python package form to give the ability to use the potentials directly in a nuclear reaction/structure code.
In addition to using Woods-Saxon/Fermi or Gaussian functions, the code also allows for the definition of nuclear matter densities using pre-calculated densities in a data file.
The manuscript provides an overview of the double folding model and the use of the code.
\end{abstract}


\end{frontmatter}



{\bf PROGRAM SUMMARY}

\begin{small}
\noindent
{\em Program Title: BiFold}                                          \\
{\em CPC Library link to program files:} (to be added by Technical Editor) \\
{\em Developer's repository link:} \url{https://github.com/mkarakoc/BiFold} \\
{\em Code Ocean capsule:} (to be added by Technical Editor)\\
{\em Licensing provisions:} GPLv3\\
{\em Programming language:} Python 3.x               \\
{\em Nature of problem:}
BiFold calculates the real part of the nuclear potential between two colliding spherical nuclei by 
integrating a density-independent/dependent nucleon-nucleon (NN) interaction  \cite{summ:SATCHLER1979183,summ:KOBOS198265,summ:KHOA200734} 
over the nuclear matter densities of the two nuclei. The code based on M3Y Reid/Paris NN interactions \cite{summ:SATCHLER1979183,summ:KOBOS198265,summ:KHOA200734}  by default, 
but it is possible to define custom NN interactions when necessary.\\
{\em Solution method:} The code uses the Fourier transform method in spherical coordinates to calculate the potential.
The method simplifies the sixfold integration \cite{summ:SATCHLER1979183} and makes the calculation a lot faster. 
The integration is done by default using Simpson's integration method, but Filon's integration method is also available.
\\

\end{small}

\section{Introduction}
\label{sec:intro}
The folded potential model \cite{SATCHLER1979183} is a well-known model for describing the mean-field nuclear interaction between two colliding nuclei. 
It has been widely used in the literature \cite{SATCHLER1979183,KOBOS198265,KHOA200734,KARAKOC200673} since accepted as a more realistic approach than a phenomenological 
potential such as the well-known Woods-Saxon. The latter usually needs three free parameters to fit data, and this may create possible known ambiguities \cite{SATCHLER1985book}.
On the contrary, a folded potential usually has only one free parameter called the re-normalization factor to compensate for higher-order effects in a nuclear reaction. Folded potentials have two categories single folded (SF) potentials and double folded (DF) potentials. 

The SF potential describes the interaction between two nuclei using one of the two nuclei's densities 
and a phenomenological nucleon-nucleus interaction potential.
The shortcomings of this treatment are density dependence, 
surface features, or couplings of a nuclear system are not well defined, as pointed out in the report \cite{SATCHLER1979183} of Satchler and Love. 
These shortcomings make the depth of the SF potentials unrealistically deep \cite{GOLDBERG1978,KOBOS198265} to explain the rainbow scattering.

The usual DF potential between two spherical nuclei is constructed by integrating over an effective nucleon-nucleon (NN) interaction with nuclear matter densities representing nucleons of both nuclei. Although, DF potentials would overcome the shortcomings of SF potentials.
Some cases for DF potentials  can still overestimate the depth of the nuclear potential. Additionally, it is necessity to add the antisymmetrization-exchange effects and the density-dependent saturation effects in the effective NN interaction \cite{KOBOS198265} for a more realistic nuclear potential.

The main reason for density dependence (DD) at NN interaction is the Pauli principle effects in the nuclear medium of both colliding nuclei. The DD of NN interaction has several treatments in the literature, but the code BiFold is built on the treatments of Satchler and Love \cite{SATCHLER1979183}, Kobos \etal \cite{KOBOS198265}, and Khoa \etal \cite{KHOA200734}. All these three treatments have the frozen density approximation while the case of Kobos \etal\cite{KOBOS198265} has energy dependence, but the other two have no energy dependence.

Many studies \emph{(see the references in the present work)} have used these DD treatments of NN interactions, 
but there are very few published codes \cite{COOK1982125,GONTCHAR2010168} to be able to reproduce the results of these works. 
In addition, the codes have limitations, and not all are updated regularly. 
For example, DFPOT \cite{COOK1982125} cannot calculate the potentials with the DDM3Y, BDM3Y$n$ ($n=1,2,3$), 
and CDM3Y$n$ ($n=1\ldots6$) type density-dependent interactions \cite{KHOA200734}.
While DFMSPH \cite{GONTCHAR2010168} can calculate many of those, it does not support BDM3Y2, BDM3Y3 \cite{KHOA19956},
and the first version of DDM3Y \cite{KOBOS198265}.
And it is a well-known fact results of a study must be reproducible in science.  
Therefore, the code BiFold will help the community in these manners.

\section{The model}
\label{sec:model}

\subsection{The usual double folded potential}
\label{sub:df}

It can be claimed that Coulomb potential between two spherical charges is the inspiration for the double folded potentials. And it is formulated as \cite{SATCHLER1979183,KARAKOC200673}
\begin{equation}
    U_{DF}(\vec R) = \int \int d\vec r_p d\vec r_t ~\rho_p(\vec r_p) \rho_t(\vec r_t) v({\vec s}),
    \label{eq:df}
\end{equation}
where $\rho_p$ and $\rho_t$ are charge densities, and $v(\vec s)=1/s$ interaction between charges in Coulomb potential case, while $\rho_p$ and $\rho_t$ are nuclear matter densities, and $v(\vec s)$ is the effective NN interaction between point-like nucleons in nuclear potential case. Its schematic representation and definitions of the vectors are given in Fig. \ref{eq:df}.

\begin{figure}
    \centering
    \includegraphics[width=0.5\textwidth]{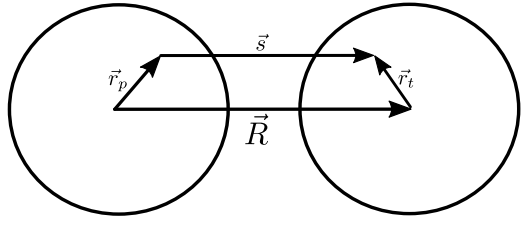}
    \caption{It is the schematic representation of DF potential in Eq.~(\ref{eq:df}) where $\vec R$ is the vector between the centers of the projectile (p) and the target (t) nuclei. $\vec r_p$ and $\vec r_t$ are the locations of the interacting parts of the nucleon distributions of both nuclei. And $\vec s = \vec R + \vec r_t - \vec r_p$ is the separation between them.}
    \label{fig:df}
\end{figure}

The effective NN interaction $v(\vec s)$ is density-independent in this usual definition of the DF potentials. The medium effects are included in the calculations by changing $v(\vec s)$ with $v(\rho, \vec s)=F(\rho) v(\vec s)$, where $F(\rho)$ describes the density dependence of the NN interaction.

\subsection{Effective density-independent NN interactions}
\label{sub:diNN}

Although BiFold can use a wide range of density-independent NN interactions in the DF potential calculations, 
M3Y type interactions \cite{BERTSCH1977399,KHOA19956,BRANDAN1997143} are defined by default in the code since they are perhaps the most widely used ones.

These interactions are called M3Y-Reid \cite{REID1968411} and M3Y-Paris \cite{LACOMBE198021} effective interactions.
The former is derived from the solution of the Bethe-Goldstone equation with Reid \cite{REID1968411} soft-core interaction 
on a harmonic oscillator basis to obtain G-matrix. The latter is derived from a more fundamental Paris NN potential \cite{LACOMBE198021}
to generate all components of the effective interaction \cite{ANANTARAMAN1983269}.

Both versions of the NN interactions have direct ($v_{d}$) and exchange ($v_{ex}$) parts,
\begin{align}
\text{M3Y-Reid:} \nonumber\\ 
\quad v_{d}(\vec s) &=\left[7999 \frac{\mathrm{e}^{-4 s}}{4 s}-2134 \frac{\mathrm{e}^{-2.5 s}}{2.5 s}\right]~\mathrm{MeV}, 
\label{eq:reidD}\\
\text{ZR:} \quad v_{ex}(\vec s) &= J_{ex}(E) \delta(\vec s) \quad \text{where} \quad J_{ex}(E) \approx-276[1-0.005 \varepsilon]~\mathrm{MeV} \mathrm{fm^{3 }},
\label{eq:reidZR}\\
\text{FR:} \quad v_{ex}(\vec s) &=\left[4631 \frac{\mathrm{e}^{-4 s}}{4 s}-1787 \frac{\mathrm{e}^{-2.5 s}}{2.5 s}-7.847 \frac{\mathrm{e}^{-0.7072 s}}{0.7072 s}\right]~\mathrm{MeV}, 
\label{eq:reidFR}
\end{align}
\begin{align}
\text{M3Y-Paris:} \nonumber\\ 
\quad v_{d}(\vec s) &=\left[11062 \frac{\mathrm{e}^{-4 s}}{4 s}-2538 \frac{\mathrm{e}^{-2.5 s}}{2.5 s}\right]~\mathrm{MeV},
\label{eq:parisD}\\
\text{ZR:} \quad v_{ex}(\vec s) &= J_{ex}(E) \delta(\vec s) \quad \text{where} \quad J_{ex}(E) \approx-590[1-0.002\varepsilon]~\mathrm{MeV} \mathrm{fm}^{3},
\label{eq:parisZR}\\
\text {FR:} \quad v_{ex}(\vec s) &=\left[-1524 \frac{\mathrm{e}^{-4 s}}{4 s}-518.8 \frac{\mathrm{e}^{-2.5 s}}{2.5 s}-7.847 \frac{\mathrm{e}^{-0.7072 s}}{0.7072 s}\right]~\mathrm{MeV}, 
\label{eq:parisFR}
\end{align}
while the direct one defines the usual nuclear interaction between nucleons.
The exchange part describes the interchange (knock-on exchange) of the nucleons of the colliding nuclei \cite{KHOA19956,BRANDAN1997143}.
The DF potential will have two parts since the effective NN interactions have two parts; then Eq.~(\ref{eq:df}) will have a new form
\begin{align}
    U_{DF}(\vec R) &= U_{D}(\vec R) + U_{EX}(\vec R) = \nonumber\\ 
    &\int \int d\vec r_p d\vec r_t ~\rho_p(\vec r_p) \rho_t(\vec r_t) v_d({\vec s}) +
    \int \int d\vec r_p d\vec r_t ~\rho_p(\vec r_p) \rho_t(\vec r_t) v_{ex}({\vec s}),
    \label{eq:df_d_ex}
\end{align}
where $U_D$ and $U_{EX}$ direct and exchange folding potentials, respectively.

There are two approaches for the exchange parts of the interactions;  these are zero-range (ZR) \cite{LOVE197574,SATCHLER199449} 
or finite-range (FR) \cite{KHOA1988376} knock-on exchange interactions. The ZR approaches given in Eqs. (\ref{eq:reidZR})
and (\ref{eq:parisZR}), where $\varepsilon = E/a_p$ is the projectile's incident energy per nucleon in the laboratory frame, are widely used in the literature due to their simplicity in calculations. 
One needs only to put the ZR interactions in  \ref{eq:reidZR} or Eq.~(\ref{eq:parisZR}) to the exchange part of 
Eq.~(\ref{eq:df_d_ex}) to obtain the exchange part of the DF potential ($U_{EX}$). The depths of the ZR interactions are defined by $J_{ex}(E)$. 
Determination of $J_{ex}(E)$ is empirically done; the detailed information can be found in Refs. \cite{LOVE197574,SATCHLER199449,BRANDAN1997143}.

As is pointed out by Khoa \cite{KHOA1988376}, the exchange interaction, in general, must be nonlocal \cite{SINHA19751}. 
Thus, the exact numerical calculation of exchange interaction can become too complicated. 
A plain wave \cite{SINHA1979289} approximation for the relative motion of nucleons can overcome this complication and lead to 
an equivalent local potential. The plain wave given in Ref. \cite{KHOA200734} is
\begin{equation}
\chi(\vec R + \vec s) \approx \exp \left(\frac{ i \vec K ( \vec R) \cdot \vec s }{M}\right) \chi( \vec R ),
\label{eq:localmom}
\end{equation}
where $M = a_p a_t/(a_p + a_t)$ is the recoil factor (or reduced mass), while $a_p$ and $a_t$ are the mass numbers of the
projectile and target nuclei, respectively. And $\vec K(\vec R)$ is the local momentum of the relative motion  given by \cite{SINHA19751}
\begin{equation}
K^{2}( \vec R )=\frac{2 m M}{\hbar^{2}}\left[E_{\text {c.m. }} - U_D(\vec R ) - U_{EX}(\vec R ) - U_{ C }(\vec R )\right],
\label{eq:rel_mom}
\end{equation}
where $E_{\text {c.m. }}$ is relative energy in the center-of-mass system, $m$ is the nucleon mass, and $U_C$ is the Coulomb potential.
Then, local exchange potential will take the form of \cite{KHOA1988376,SINHA19751,BRIEVA1977299,BRIEVA1977317,CHAUDHURI1985415,CHAUDHURI1986169}:
\begin{align}
    U_{EX}(\vec R) = \int \int d\vec r_p d\vec r_t ~\rho_p(\vec r_p, \vec r_p + \vec s) \rho_t(\vec r_t, \vec r_t - \vec s) v_{ex}({\vec s}) \exp \left(\frac{ i \vec K ( \vec R) \cdot \vec s }{M}\right),
    \label{eq:df_ex}
\end{align}
where $\rho_p(\vec r_p, \vec r_p + \vec s)$ and $\rho_t(\vec r_t, \vec r_t - \vec s)$ are one-body density matrices 
\cite{CHAUDHURI1985415,CHAUDHURI1986169,GUPTA19841093} of the projectile and target nucleons. This local potential 
becomes an FR exchange potential when $v_{ex}(\vec s)$ is chosen as one of the FR interactions 
in Eqs. (\ref{eq:reidFR}) or (\ref{eq:parisFR}) (M3Y-Reid/Paris-FR). One should realize that the exchange part 
of Eq.~(\ref{eq:df_d_ex}) needs to be replaced by Eq.~(\ref{eq:df_ex}) for the FR exchange potential.

The next step in the exchange potential (Eq.~(\ref{eq:df_d_ex})) is the calculation of density matrices. 
Although the matrices can be obtained from single-particle wave functions \cite{SINHA19751}, Khoa \cite{KHOA1988376}
has chosen a realistic local approximation from Ref. \cite{CAMPI1978263}:
\begin{equation}
    \rho_{p,t}(\vec r, \vec r \pm \vec s) \approx \rho_{p,t} \left(\vec r \pm \frac{\vec s}{2}\right) ~ \hat j_1 \left (k_{\text{F}_{p,t}}(\vec r \pm \frac{\vec s}{2}) s \right),
\end{equation}
where $\hat{j_{1}}(x)=3(\sin x-x \cos x) / x^{3}$. $k_F$ is the average local Fermi momentum from Refs. \cite{CAMPI1978263,RINGSCHUCK1980,KHOA199449,KHOA2001034007}:
\begin{equation}
k_{F}(\vec r)=\left\{\left[\frac{3}{2} \pi^{2} \rho(\vec r)\right]^{2 / 3} + C_{S}  \frac{5}{3} \frac{[\nabla \rho(\vec r)]^{2}}{\rho^{2}(\vec r)}
+\frac{5}{36} \frac{\nabla^{2} \rho(\vec r)}{\rho(\vec r)}\right\}^{1 / 2},
\end{equation}
where rho is the nuclear matter densities of the projectile or the target and $C_S$ is the strength of the Weizsäcker term, representing the surface contribution to the kinetic energy density\cite{KHOA199449}.
The strength term is usually $C_S \approx \frac{1}{36}$ in the literature, but Khoa \emph{\etal} \cite{KHOA199449} have taken it as 
$C_S \approx \frac{1}{4}$ for the given reasons in their work. The default value is $C_S \approx \frac{1}{36}$ in BiFold, but the user has the option to change the value.

After this point, the FR exchange potential will have the following form \cite{KHOA1988376}:
\begin{equation}
U_{ EX }( \vec R )=4 \pi \int_{0}^{\infty} v_{ EX }(s) s^{2} d s \int f_{p}( \vec r , \vec s) f_{t}( \vec r - \vec R , \vec s) j_{0}(K( \vec R ) s / M ) d \vec r
\end{equation}
where $f_{p,t}(\vec r, \vec s) = \rho_{p,t} (\vec r) ~ \hat j_1 (k_{\text{F}_{p,t}}(\vec r) s )$ and $j_{0}(x)=\sin x / x$. Now, the exchange potential with Fourier transforms in spherical coordinates will take the form \cite{SATCHLER1979183,KHOA1988376}:
\begin{equation}
U_{ EX }( R )=4 \pi \int_{0}^{\infty} G(R,s) j_{0}(K( R ) s / M ) v_{ EX }(s) s^{2} ds,
\end{equation}
where
\begin{equation}
G(R, s)=\frac{1}{2 \pi^{2}} \int_{0}^{\infty} f_{p}(q, s) f_{t}(q, s) j_{0}(q  R) q^{2} dq,
\end{equation}

\begin{equation}
f_{p,t}(q, s)=4 \pi \int_{0}^{\infty} f_{p,t}(r, s) j_{0}(q r) r^{2} dr.
\end{equation}

Finally, it is necessary to solve a self-consistency problem to obtain the exchange part (Eq.~(\ref{eq:df_ex})) of the double folded potential (Eq.~(\ref{eq:df_d_ex})) at each radial point. 
Since the exchange potential (Eq.~(\ref{eq:df_ex})) contains the local momentum of the relative motion (Eq.~(\ref{eq:rel_mom})) and $\vec K(\vec R)$ depends on the total double folded potential, 
this problem can be solved exactly by an iterative method given in Refs. \cite{KHOA199449,KHOA199756,CHAUDHURI1985415,CHAUDHURI1986169}.

\subsection{Effective density-dependent NN interactions}
\label{sub:ddNN}

The effective density-dependent NN interaction is proposed \cite{KOBOS198265,KHOA199449,KHOA2001034007,KHOA20003} in the following form for 
both direct ($v_d$) and exchange ($v_{ex}$) parts;
\begin{equation}
\label{eq:vdde}
    v_{d,ex}(\rho, E, \vec s) = g(E) F_{d,ex}(\rho) v_{d,ex}(\vec s)
\end{equation}
where $\rho$ is the overlapping density of the nuclear medium of both nuclei.
$g(E)$ is the weak intrinsic energy dependence proposed by Khoa \etal \cite{KHOA19938}.
The density-dependent folding potential with an FR exchange part can be calculated by replacing this new form with
$v_d(\vec s)$ in Eq.~(\ref{eq:df_d_ex}) and with $v_{ex}(\vec s)$  in Eq.~(\ref{eq:df_ex}). 
In the case of a ZR exchange part \cite{HAGINO200674}, it can be calculated by using the new form for both $v_d(\vec s)$ and $v_{ex}(\vec s)$ in Eq.~(\ref{eq:df_d_ex}).

The overlapping density for the direct part and the ZR exchange part of the folded potential has been approximated in 
most of the folding potential calculations \cite{KOBOS198265,KOBOS1984205,FARID1985525,KHOA199449,KHOA19956,KHOA199574,KHOA199756,HAGINO200674} as
\begin{equation}
    \rho = \rho_p(\vec r_p) + \rho_t(\vec r_t),
\end{equation}
since it permits the separation of variables in the integrals of Eqs. (\ref{eq:df_d_ex}) and (\ref{eq:df_ex}). The overlapping density 
for the FR exchange part of the folded potential has been assumed \cite{KOBOS198265,KOBOS1984205,FARID1985525,KHOA199449,KHOA19956,KHOA199574,KHOA199756,HAGINO200674} as
\begin{equation}
    \rho = \rho_p(\vec r_p + \frac{\vec s}{2}) + \rho_t(\vec r_t - \frac{\vec s}{2}).
\end{equation}

\begin{table}
\begin{small}
\caption{\label{tab:interactions}BiFold can calculate double folding potentials for the interactions marked with ``$\checkmark$'' 
where ZR and FR stand for zero-range and finite-range exchange interactions, respectively. 
The {\Large$\times$} stands for not supported or non existing interactions.
The nuclear incompressibility values ($K[\mathrm{MeV}]$) are only exist for finite range versions of the interactions \cite{KHOA199574,KHOA199756}.}
\begin{adjustbox}{width=\textwidth}
\begin{tabular}{lllllllSlcl} 
	\hline\hline
	\multicolumn{2}{c}{Interaction names} & ZR           & FR             &   $C$   & $\alpha$ 	  & $\beta~[\mathrm{fm}^{3}]$ & $\gamma[\mathrm{fm}^{3n}]$ & n &$K[\mathrm{MeV}]$ & Refs. \\ \hline
	DDM3Y      &  Reid & $\checkmark$ & \Large$\times$  & $C(E)$  & $\alpha(E)$   & $\beta(E)$ &   0.0  &  0  &     &\cite{KOBOS198265,FARID1985525} \\ \hline
	DDM3Y1     & Reid  &$\checkmark$ & $\checkmark$  & 0.2845  & 3.6391 	     & 2.9605	   &  0.0   & 0   & 171 &\cite{KHOA19938,KHOA199449,KHOA19956,KHOA200734} \\
	DDM3Y1     & Paris & $\checkmark$ & $\checkmark$  & 0.2963  & 3.7231 	     & 3.7384	   &  0.0   & 0   & 176 & \cite{KHOA19956,KHOA199574,KHOA200734}			 \\ \hline
	BDM3Y0     & Reid  &\Large$\times$&\Large$\times$ & 1.3827  & 0.0            & 0.0        & 1.1135 & 2/3 & 232 & \cite{KHOA19938}						 \\ \hline
	BDM3Y1     & \multirow{3}{*}{Reid}      & $\checkmark$ & $\checkmark$  & 1.2253  & 0.0            & 0.0        & 1.5124 & 1   & 232 &										 \\
	BDM3Y2     &       & $\checkmark$ & $\checkmark$  & 1.0678  & 0.0            & 0.0        & 5.1069 & 2   & 354 & \cite{KHOA19938,KHOA199449,KHOA19956,KHOA200734}\\
	BDM3Y3     &       & $\checkmark$ & $\checkmark$  & 1.0153  & 0.0            & 0.0        & 21.073 & 3   & 475 &										 \\ \hline
	BDM3Y1     & \multirow{3}{*}{Paris}      & $\checkmark$ & $\checkmark$  & 1.2521  & 0.0            & 0.0        & 1.7452 & 1   & 270 &										 \\
	BDM3Y2     &       & $\checkmark$ & $\checkmark$  & 1.0664  & 0.0            & 0.0        & 6.0296 & 2   & 418 & \cite{KHOA19956,KHOA199574,KHOA200734}			 \\
	BDM3Y3     &       & $\checkmark$ & $\checkmark$  & 1.0045  & 0.0            & 0.0        & 25.115 & 3   & 566 &										 \\ \hline
	CDM3Y1     & \multirow{6}{*}{Paris}      & $\checkmark$ & $\checkmark$  & 0.3429  & 3.0232 	     & 3.5512 	   & 0.5    & 1   & 188 & \multirow{6}{*}{\cite{KHOA199756,KHOA200734}}	 \\
	CDM3Y2     &       & $\checkmark$ & $\checkmark$  & 0.3346  & 3.0357 	     & 3.0685 	   & 1.0    & 1   & 204 &										 \\
	CDM3Y3     &       & $\checkmark$ & $\checkmark$  & 0.2985  & 3.4528 	     & 2.6388 	   & 1.5    & 1   & 217 &										 \\
	CDM3Y4     &       & $\checkmark$ & $\checkmark$  & 0.3052  & 3.2998 	     & 2.3180 	   & 2.0    & 1   & 228 &										 \\
	CDM3Y5     &       & $\checkmark$ & $\checkmark$  & 0.2728  & 3.7367 	     & 1.8294 	   & 3.0    & 1   & 241 &										 \\
	CDM3Y6     &       & $\checkmark$ & $\checkmark$  & 0.2658  & 3.8033 	     & 1.4099 	   & 4.0    & 1   & 252 &										 \\ \hline
	M3Y        & Reid  & $\checkmark$ & $\checkmark$  & \multicolumn{6}{c}{\multirow{2}{*}{usual density-independent M3Y}}    & \multirow{2}{*}{\cite{SATCHLER1979183,KHOA1988376,HAGINO200674}}   \\
    M3Y        & Paris & $\checkmark$ & $\checkmark$  &         &  	             & 	       &        &     &     &                 \\ 
	\hline\hline
\end{tabular}
\end{adjustbox}
\end{small}
\end{table}

The code can calculate double folding potentials for the interactions marked with a ``$\checkmark$'' in Table \ref{tab:interactions}.
The density dependence of these interactions is defined by $F(\rho)$ in Eq.~(\ref{eq:vdde}). Different versions of \ref{eq:vdde} are proposed in the Refs. \cite{KOBOS198265,KHOA19938,KHOA19956,KHOA199756}.
In a more recent study by Khoa \etal \cite{KHOA200734}, these different versions of $F(\rho)$ merged into one formula;
\begin{equation}
	\label{eq:Frho}	
	F(\rho)=C[1+\alpha \exp (-\beta \rho)-\gamma \rho^{n}].
\end{equation}
The parameters of this formula are given in Table \ref{tab:interactions} other than the original DDM3Y \cite{KOBOS198265} since its parameters are energy-dependent.
The values of these energy-dependent parameters can be obtained from the Refs. \cite{KOBOS198265,FARID1985525}.

The final part of the density-dependent NN interaction (Eq.~(\ref{eq:vdde})) is $g(E)$. 
It is $g(E) = 1$ for the original DDM3Y \cite{KOBOS198265} since $g(E)$ does not exist for this interaction.
For the remaining density-dependent interactions in Table \ref{tab:interactions}, 
$g(E) \approx 1 - \kappa \varepsilon$, where $\kappa=0.002$ and $\kappa=0.003$ for the M3Y-Reid and M3Y-Paris types of interactions, respectively, 
and $\varepsilon = E/a_p$ is energy (in MeV) per nucleon.

\section{The code}
\label{sec:code}
\begin{figure}
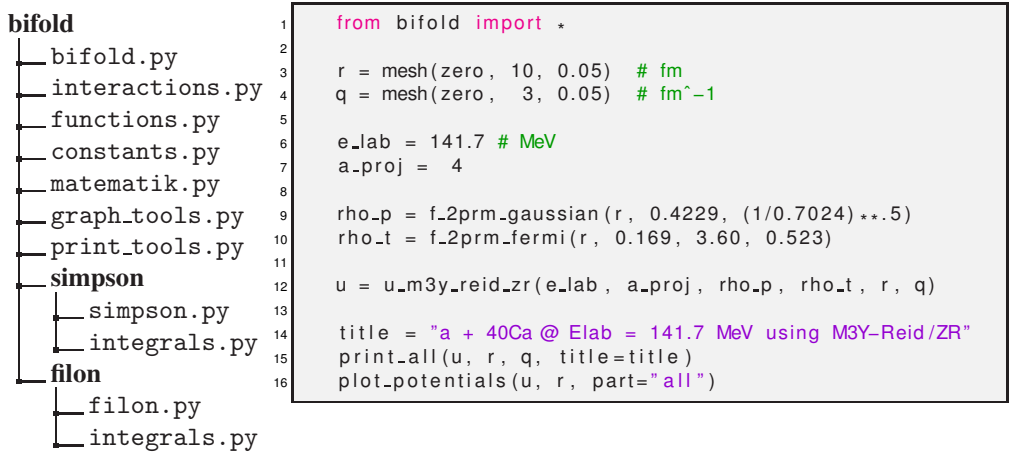

	\centering
	\begin{minipage}{.31\textwidth}
		\dirtree{
			.1 \bf{bifold}.
			.2 bifold.py.
			.2 interactions.py.
			.2 functions.py.
			.2 constants.py.
			.2 matematik.py.
			.2 graph\_tools.py.
			.2 print\_tools.py.
			.2 \bf{simpson}.
			.3 simpson.py.
			.3 integrals.py.
			.2 \bf{filon}.
			.3 filon.py.
			.3 integrals.py.
		}
	\end{minipage}
	\begin{minipage}{.68\textwidth}
		\begin{lstlisting}
			from bifold import *
			
			r = mesh(zero, 10, 0.05)  # fm
			q = mesh(zero,  3, 0.05)  # fm^-1
			
			e_lab = 141.7 # MeV
			a_proj =  4
			
			rho_p = f_2prm_gaussian(r, 0.4229, (1/0.7024)**.5)
			rho_t = f_2prm_fermi(r, 0.169, 3.60, 0.523)
			
			u = u_m3y_reid_zr(e_lab, a_proj, rho_p, rho_t, r, q)
			
			title = "a + 40Ca @ Elab = 141.7 MeV using M3Y-Reid/ZR"
			print_all(u, r, q, title=title)
			plot_potentials(u, r, part="all")
		\end{lstlisting}
	\end{minipage}
	\caption{The file tree structure of BiFold code is on the left. The bold ones are directories, and the rest are Python files. 
		And an example Python file to calculate a double folding potential for an $\alpha + ^{40}$Ca elastic scattering on the right.}
	\label{fig:bifold}
\end{figure}

The file structure of BiFold code given is in Fig. \ref{fig:bifold} on the left.
The bold ones are directories, and the rest are Python files in the given file structure.
The two sub-directories in the {\bf bifold} directory ({\bf simpson} and {\bf filon}) 
have the same structures other than the integration methods.
\bifoldpy\ file calls \simpsonpy\ file since the Simpson integration method is
the default integration method of BiFold. Both \simpsonpy\ and \filonpy\ have the functions with 
the names starts-with \postem{u\_} since the symbols of the double folding potentials are $U_{DF}$ or
$U_{D}$ or $U_{EX}$ in Eq.~(\ref{eq:df_d_ex}). Therefore, \simpsonpy\ (\filonpy) is
the main part of the code where the calculation of the double folding integrals takes place 
through \postem{integrals.py}.
Table \ref{tab:dfpots} shows the names of these functions that correspond to the calculation of 
double folding potential types. Table \ref{tab:dfpots} also lists the Python functions of the \postem{interactions.py} 
file for the computation of the effective NN interactions in Eqs.~(\ref{eq:reidD}) through (\ref{eq:parisFR}).
The \postem{functions.py} file contains Python functions for nuclear matter densities and effective NN 
interactions. Table \ref{tab:functions} shows the mathematical representations of the densities and interactions 
versus Python representations. The \postem{matematik.py} contains the spherical Bessel functions, 
numerical derivation functions, and other mathematical tools. The \postem{constants.py} is for all the necessary
physical and mathematical constants. The other two files (\postem{graph\_tools.py} and \postem{print\_tools.py})
are for drawing and printing the results.
\begin{table}[!hbtp]
	\begin{small}
\caption{\label{tab:dfpots} The presented names are the Python functions in BiFold 
	to calculate the effective NN interactions and the doubling potentials.
	The {\Large$\times$} stands for not supported or non existing potentials/interactions.}
		\begin{adjustbox}{width=\textwidth}
			\begin{tabular}{ll|ll|lll} 
				\hline\hline
            	\multicolumn{2}{c|}{Interaction names} & \multicolumn{2}{c|}{\simpsonpy / \filonpy}             &   \multicolumn{3}{c}{\postem{interactions.py}}  \\
                                                    &  & $U_{DF}$ with ZR          & $U_{DF}$ with FR             &   $v_d(\vec s)$   & ZR: $v_{ex}(\vec s)$ 	  & FR: $v_{ex}(\vec s)$ \\ \hline
				DDM3Y      &  Reid & u\_ddm3y\_reid\_zr & \multicolumn{1}{c|}{\Large$\times$}  & v\_m3y\_reid\_d  & v\_m3y\_reid\_ex\_zr   & \multicolumn{1}{c}{\Large$\times$}  \\ \hline
				DDM3Y1     & \multirow{2}{*}{Reid}      & \multirow{2}{*}{u\_xdm3yn\_zr} & \multirow{2}{*}{u\_xdm3yn\_fr}  & \multirow{2}{*}{v\_m3y\_reid\_d}  & \multirow{2}{*}{v\_m3y\_reid\_ex\_zr}   & \multirow{2}{*}{v\_m3y\_reid\_ex\_fr}   \\
				BDM3Y$n$; $n=1,2,3$     &       &  &   &   &   &   \\ \hline
				DDM3Y1     & \multirow{3}{*}{Paris}      & \multirow{3}{*}{u\_xdm3yn\_zr} & \multirow{3}{*}{u\_xdm3yn\_fr}  & \multirow{3}{*}{v\_m3y\_paris\_d}  & \multirow{3}{*}{v\_m3y\_paris\_ex\_zr}   & \multirow{3}{*}{v\_m3y\_paris\_ex\_fr}   \\
				BDM3Y$n$; $n=1,2,3$     &       &  &   &   &   &   \\ 
				CDM3Y$n$; $n=1\ldots6$ &       &  &   &   &   &   \\ \hline 
				M3Y        & Reid  & u\_m3y\_reid\_zr &   \multirow{2}{*}{u\_xdm3yn\_fr}  & v\_m3y\_reid\_d  & v\_m3y\_reid\_ex\_zr   & v\_m3y\_reid\_ex\_fr  \\ 
				M3Y        & Paris & u\_m3y\_paris\_zr &                                  &	v\_m3y\_paris\_d & v\_m3y\_paris\_ex\_zr  & v\_m3y\_paris\_ex\_fr  \\ 
				\hline\hline
			\end{tabular}
		\end{adjustbox}
	\end{small}
\end{table}

\begin{table}[!hbtp]
	\begin{small}
		\begin{adjustbox}{width=\textwidth}		
			\begin{threeparttable}[b]			
				\caption{\label{tab:functions}	The presented names are the Python functions in BiFold to calculate 
					the nuclear matter densities or, if necessary, to write a new type of interaction or 
					to calculate phenomenological potentials.}
				\begin{tabular}{ll|ll} 
					\hline\hline
					Python function & Math formula & 	Python function & Math formula \\ \hline
					f\_exp\_decay(r, V0, a)        & $V_0~\textrm{e}^{-a r}$                              & f\_sog(r, Ris, Qis, RP, Ze)             &  $\sum\limits_{i} A_i\left\{\textrm{e}^{-\left[(r-R_i) / \gamma\right]^2}+\textrm{e}^{-\left[(r+R_i) / \gamma\right]^2}\right\}$  \cite{DEVRIES1987495}   \\ 
                    f\_yukawa(r, V0, a, b)         & $V_0~\textrm{e}^{-a r}/b r$                          & f\_dirac\_delta(r, V0)                  &  $V_0~\delta(r)$ \\
                    f\_2prm\_fermi(r, V0, R, a)    & $V_0~\frac{1}{1 + \textrm{e}^{\frac{r-R}{a}}}$       & f\_external(r, external\_data)\tnote{1} &  reads from a local file \\
                    f\_3prm\_fermi(r, V0, w, R, a) & $V_0~\frac{1+w r^2}{1 + \textrm{e}^{\frac{r-R}{a}}}$ & f\_internet(r, url)\tnote{1}            &  reads a file from a given url \\
                    f\_2prm\_gaussian(r, V0, a)    & $V_0~\textrm{e}^{-(r/a)^2}$                          & f\_ripl(r, Z, A)\tnote{1}               &  HFB14 densities from RIPL \cite{CAPOTE20093107,GORIELY2007064312,AUDI2003337,ANGELI2004185}\\
                    f\_3prm\_gaussian(r, V0, w, a) & $V_0~(1+w r^2)~\textrm{e}^{-(r/a)^2}$                &                                         &  \url{https://www-nds.iaea.org/RIPL} \\					
					\hline\hline
				\end{tabular}
				\begin{tablenotes}
					\item [1] These functions create Python arrays using data from local or remote files. 
					These arrays contain interpolated values to match the given $r$ array defined by the $mesh(r_{min}, r_{max}, dr)$ function.
				\end{tablenotes}		
			\end{threeparttable}	
		\end{adjustbox}
	\end{small}
\end{table}

\begin{figure}[!hbtp]
	\centering
	\includegraphics[width=0.6\textwidth]{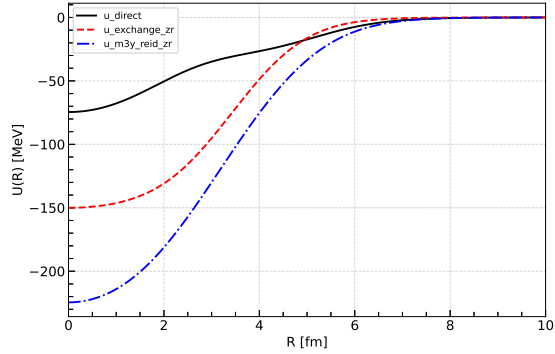}
	\caption{It is the double folding potential [dash-dot] with direct [solid] and exchange [dash] parts for the $\alpha + ^{40}$Ca elastic scattering.}
	\label{fig:m3y_reid_zr}
\end{figure}

An example Python file for calculating a density-independent double potential using M3Y-Reid NN interaction with the ZR exchange part (see Eqs. (\ref{eq:reidD}) and (\ref{eq:reidZR})) 
for an $\alpha + ^{40}$Ca elastic scattering is on the right side of Fig. \ref{fig:bifold}.
And the results of the calculation are shown in Figs. \ref{fig:m3y_reid_zr_output} and \ref{fig:m3y_reid_zr}. One may agree that the code is easy to understand.
The first line of the Python code imports BiFold since it is a Python package.
On the third and fourth lines, the \emph{mesh} functions defines the numerical integration grids for
$r=zero$~fm to $r=10$~fm with $0.05$~fm steps and 
$q=zero$~fm$^{-1}$ to $r=3$~fm$^{-1}$ with $0.05$~fm$^{-1}$ steps where $zero=1\times10^{-10}$.
The sixth and seventh lines are the laboratory energy (\emph{e\_lab}) and the atomic mass number (\emph{a\_proj}) of 
the projectile ($\alpha~-~\textrm{particle}$), respectively.
The ninth and tenth lines are the nuclear matter densities of the projectile (\emph{rho\_p}) and the target (\emph{rho\_t}) nuclei, respectively. 
The twelfth line is the double folding calculation in Eq.~(\ref{eq:df_d_ex}) for M3Y-Reid NN interaction with the ZR exchange part (see Eqs. (\ref{eq:reidD}) and (\ref{eq:reidZR})).
The rest of the file is optional if one needs to print (Fig. \ref{fig:m3y_reid_zr_output}) the calculation information and draw (Fig. \ref{fig:m3y_reid_zr}) the potentials versus radial distance between the two nuclei.
This code is a simple example of how to use BiFold to calculate a double-folded potential. 
The related GitHub (\url{https://github.com/mkarakoc/BiFold}) page has the BiFold code, 
a more detailed user guide, and more examples.

\begin{figure}[!hbtp]
	\centering
\begin{Verbatim}[fontsize=\fontsize{5.5}{6.3}]
	═══════════════════════════════╣a + 40Ca @ Elab = 141.7 MeV using M3Y-Reid/ZR╠═══════════════════════════════
	
	density/interaction               L          norm        renorm      vol2        vol4            msr
	──────────────────────────────────────────────────────────────────────────────────────────────────────────────
	───────total     - - - - - - - - - - - - - - - - - - - - - - - - - - - - - - - - - - - - - - - - - - - - - - - 
	u_R      : u_m3y_reid_zr          0          None        1.000   -59536.370  -982867.456       16.509 
	───────direct    - - - - - - - - - - - - - - - - - - - - - - - - - - - - - - - - - - - - - - - - - - - - - - - 
	u_R      : u_direct               0          None        1.000   -23281.699  -486747.684       20.907 
	rho_p    : f_2prm_gaussian        0          None        1.000        4.000        8.543        2.136 
	rho_t    : f_2prm_fermi           0          None        1.000       39.908      461.090       11.554 
	vnn      : f_yukawa               0          None        1.000     1570.558      588.970        0.375 
	vnn      : f_yukawa               0          None        1.000    -1716.459    -1647.805        0.960 
	───────exchange  - - - - - - - - - - - - - - - - - - - - - - - - - - - - - - - - - - - - - - - - - - - - - - - 
	u_R      : u_exchange_zr          0          None        1.000   -36254.670  -496119.772       13.684 
	rho_p    : f_2prm_gaussian        0          None        1.000        4.000        8.543        2.136 
	rho_t    : f_2prm_fermi           0          None        1.000       39.908      461.090       11.554 
	vnn      : f_dirac_delta          0          None        1.000     -227.114        0.000        0.000 
	
	  R            u_R       	   R            u_R       	   R            u_R       	   R            u_R       	
	───────   ───────────────	 ───────   ───────────────	 ───────   ───────────────	 ───────   ───────────────	
	  0.000    -224.563778558	   2.500    -157.142066726	   5.000     -33.591208578	   7.500      -1.294809340	
	  0.050    -224.538322266	   2.550    -154.527402817	   5.050     -32.017487683	   7.550      -1.195282003	
	  0.100    -224.461928239	   2.600    -151.883273803	   5.100     -30.494524998	   7.600      -1.102914652	
	  0.150    -224.334521330	   2.650    -149.211853484	   5.150     -29.022094646	   7.650      -1.017241399	
	  ...
	  ...
	  ...
	  2.300    -167.264509196	   4.800     -40.395194787	   7.300      -1.774804137	   9.800      -0.024215666	
	  2.350    -164.788335884	   4.850     -38.617900517	   7.350      -1.641422081	   9.850      -0.022007393	
	  2.400    -162.274567943	   4.900     -36.891404224	   7.400      -1.517343866	   9.900      -0.019983036	
	  2.450    -159.725143549	   4.950     -35.215829117	   7.450      -1.401989622	   9.950      -0.018129460	
\end{Verbatim}
	\caption{It is the output of the double folding potential calculation for the $\alpha + ^{40}$Ca elastic scattering.}
	\label{fig:m3y_reid_zr_output}
\end{figure}

The output of the BiFold calculation shown in Fig. \ref{fig:m3y_reid_zr_output}  gives individual information
about the potentials and the functions used in the calculations. This individual information from left to right contains 
a Python dictionary key, name of the function/potential, multi-polarity (L), normalization (norm) and re-normalization (renorm), 
volume integrals (vol2 and vol4), and mean square radii (msr). The output also contains printout of the calculated potential $U_R$
versus radial distance $R$ between two nuclei. 

The dictionary keys \emph{func\_i}, \emph{func\_r} and \emph{func\_q} 
store the information about the function, 
numeric values of the function at every point defined by $mesh(r_{min}, r_{max}, dr)$, and 
numeric values of the function's Fourier transform  at every point defined by $mesh(q_{min}, q_{max}, dq)$, respectively.
The output does not list these keys. The keys \emph{total}, \emph{direct} and \emph{exchange} are to reach total, direct 
and exchange individual parts of the calculations under the previous keys. And \emph{u\_R}, \emph{rho\_p}, \emph{rho\_t}, 
and \emph{vnn} keys store information about every individual part under the previous keys.  
Finally, one can reach the information tabulated in the output by using \emph{L}, \emph{norm}, \emph{renorm}, 
\emph{vol2}, \emph{vol4} and \emph{msr} keys. For example, using \emph{func\_i}, \emph{direct} and \emph{rho\_p}
keys in the given order will produce the following output about the density of the projectile nucleus 
in the example code in Fig. \ref{fig:bifold}:

\vspace{6pt}

\begin{center}
\begin{minipage}{.83\textwidth}
\begin{lstlisting}
input : u['func_i']['direct']['rho_p']
output: [{'name': 'f_2prm_gaussian', 'L': 0, 'norm': None, 'renorm': 1.0,
          'vol2': 4.000237160, 'vol4': 8.542647693, 'msr': 2.135535308}]
\end{lstlisting}
\end{minipage}
\end{center}

\vspace{-16pt}

The following formula defines the volume integrals for all radial dependent functions and potentials,
\begin{equation}
vol(n+2) = renorm ~ 4\pi \int f(r) r^{n+2} dr.
\end{equation}
If the value of \emph{norm} is \emph{None}, then \emph{renorm = 1}, otherwise if \emph{norm} is a real number, 
then \emph{renorm} is,
\begin{equation}
	renorm = norm  \left[ 4\pi\int f(r) r^{n+2} dr \right]^{-1}.
\end{equation}
The mean square radii (msr) of  all radial dependent functions and potentials are $\langle r^2 \rangle~=~vol4/vol2$.

\section{Test cases}

This section compares BiFold computations to three examples to demonstrate the code's reliability.
The first one is an analytical calculation, and the second one is a numerical calculation using DFPOT \cite{COOK1982125}.
And the work of Khoa \etal \cite{KHOA199574} is the last one. One can assume that the examples are reasonably accurate since the first example 
is an exact solution to the double folding integrals, the second example 
is a published code used and tested many times in the literature, 
and the final one is a reliable published work. 
Therefore, these examples are reference calculations to use in the following formula,
\begin{equation}
	\xi^2 = \frac{1}{N} \sum\limits_{i}^{N} \left(\frac{U_A(R_i)-U_B(R_i)}{U_A(R_i)+U_B(R_i)}\right)^2.
	\label{eq:mre}
\end{equation}

This formula defines a mean relative error (mre) \cite{GONTCHAR2010168}  for comparing BiFold calculations with 
reference calculations where $U_A$ and $U_B$ are the reference and the present 
double folding potentials with the radial distance $R_i$, respectively.
It is better if $\xi^2$ is getting closer to zero, as it will mean that the results 
of both computations are getting more consistent. The results of the mre calculations are in Table \ref{tab:mre}, 
and the details of the test cases are in the following sections.

\begin{table}[!hbtp]
	\center	
	\begin{small}
		\caption{\label{tab:mre} The mean relative errors, defined by Eq.~(\ref{eq:mre}), are given for the three test cases. 
			The cases are compared with BiFold's calculations using both Simpson's and Filon's integration.}
		\begin{adjustbox}{width=.9\textwidth}
			\begin{tabular}{llllll} 
				\hline\hline
				Integration   & \multicolumn{1}{c}{\multirow{2}{*}{$\xi^2$}}  & \multicolumn{2}{c}{$\overbrace{\rule{120pt}{0pt}}^{\textrm{\normalsize Analytical}}$}         & \multirow{2}{*}{DFPOT} &  \multirow{2}{*}{Khoa \etal \cite{KHOA199574}}     \\ 
				method of BiFold                 &                 & $q_{max}=3~\textrm{fm}^{-1}$  & $q_{max}=10~\textrm{fm}^{-1}$     &               &                \\ \hline				
										&  total          & $2.16\times10^{-3}$           & $6.66\times10^{-8}$ &  $1.31\times10^{-7}$ & $6.58\times10^{-5}$   \\ 				
				Simpson                 &  direct         & $1.84\times10^{2}$            & $1.19\times10^{-7}$ &  $8.10\times10^{-5}$ & $6.65\times10^{-3}$   \\ 		
										&  exchange       & $1.22\times10^{0}$            & $5.87\times10^{-6}$ &  $1.40\times10^{-6}$ & $2.48\times10^{-5}$   \\ \hline
										&  total          & $2.16\times10^{-3}$           & $6.64\times10^{-8}$ &  $1.31\times10^{-7}$ & $6.57\times10^{-5}$   \\ 				
				Filon      	   		    &  direct         & $1.84\times10^{2}$            & $1.16\times10^{-7}$ &  $8.13\times10^{-5}$ & $6.65\times10^{-3}$   \\ 		
						    			&  exchange       & $1.22\times10^{0}$            & $5.87\times10^{-6}$ &  $1.40\times10^{-6}$ & $2.48\times10^{-5}$   \\ 				
				\hline\hline
			\end{tabular}
		\end{adjustbox}
	\end{small}
\end{table}

\subsection{BiFold vs. analytical calculation}

In this case, $\alpha+\alpha$ scattering with projectile energy 50~MeV in the laboratory system is in consideration. 
Satchler and Love \cite{SATCHLER1979183} suggested a Gaussian shaped nuclear matter distribution for
an $\alpha~-~\textrm{particle}$ is
\begin{equation}
	\rho(r) = 0.4229~\textrm{e}^{-0.7024 r^2}~\textrm{fm}^{-3},
	\label{eq:gauss}
\end{equation}
with a mean square radius $\langle r^2\rangle = 2.1355~\textrm{fm}^2$. A new nuclear matter distribution
for the $\alpha~-~\textrm{particle}$ is
\begin{equation}
	\rho(r) = 2.12~\textrm{e}^{-2.3705 r}~\textrm{fm}^{-3},
	\label{eq:expdecay}
\end{equation}
proposed using the msr value since the previous one does not allow to obtain an analytical solution. 
Therefore, the reason for choosing this distribution is to obtain an exact analytical double-folded potential 
since the density-independent NN effective interaction (M3Y-Reid, Eq.~(\ref{eq:reidD}) also has a similar mathematical form. 
Then this is easily achieved by using the Fourier transform techniques as usual \cite{SATCHLER1979183}
for the double folding integral given in Eq.~(\ref{eq:df_d_ex}), 
but this time with an analytical integration. Thus, the analytical double-folded potential 
with both direct and ZR exchange NN interactions included is
\begin{align}
	U(R) =~& \frac{1}{R}~2747 \left(\textrm{e}^{-4 R}-31323.3246~\textrm{e}^{-2.5 R} -10.729866914~\textrm{e}^{-2.3705 R}~\times\right.  \nonumber\\
	& \left. [R^3 -24.345 R^2 + 377.89705 R -2919.17177]\right).
\end{align}

\begin{figure}[!hbtp]
	\centering
	\begin{subfigure}[b]{0.495\textwidth}
		\centering
		\includegraphics[width=\textwidth]{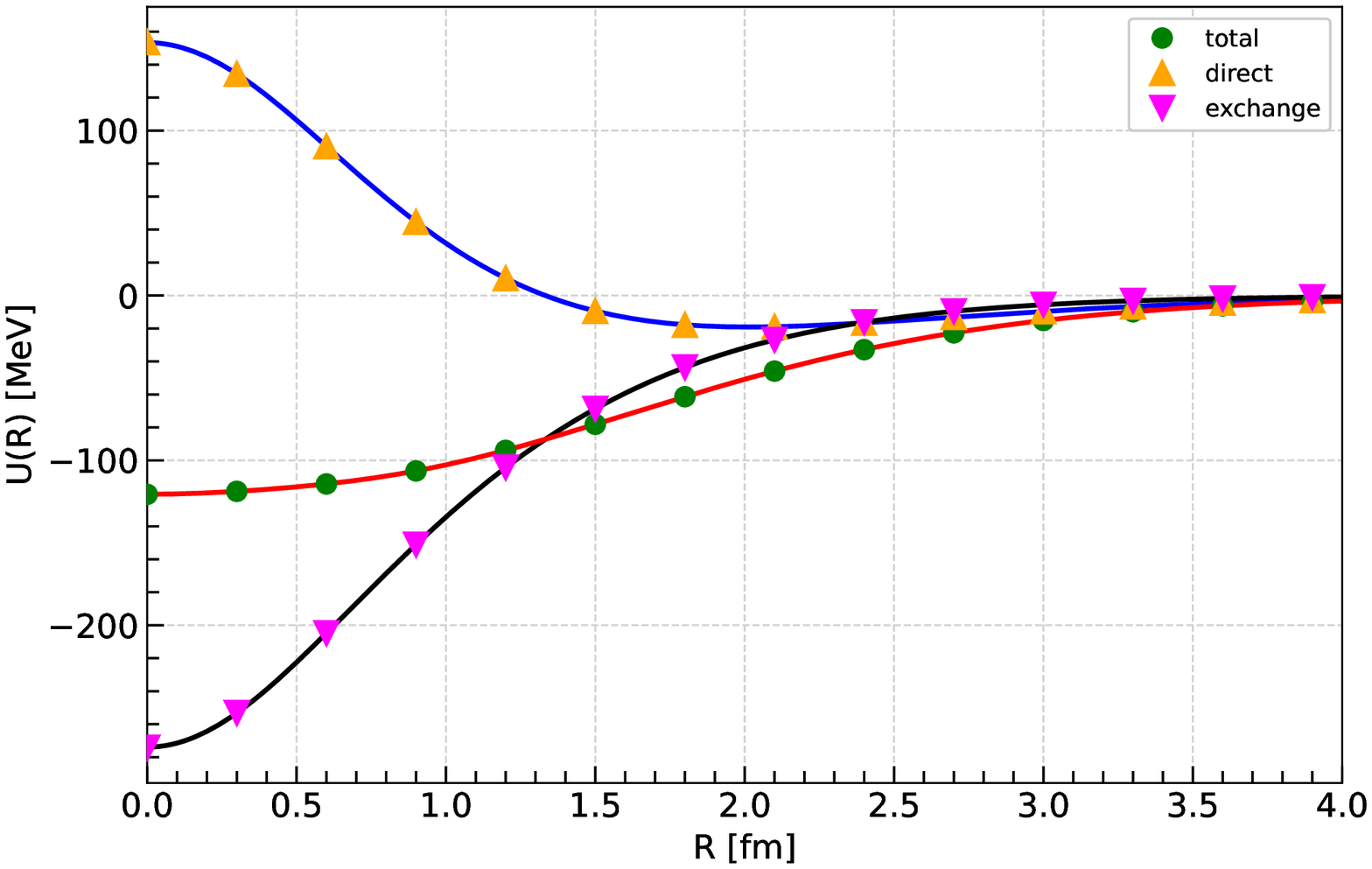}
		\caption{$q_{max}~=~10~\textrm{fm}^{-1}$}
		\label{fig:qa}
	\end{subfigure}
	\hfill
	\begin{subfigure}[b]{0.495\textwidth}
		\centering
		\includegraphics[width=\textwidth]{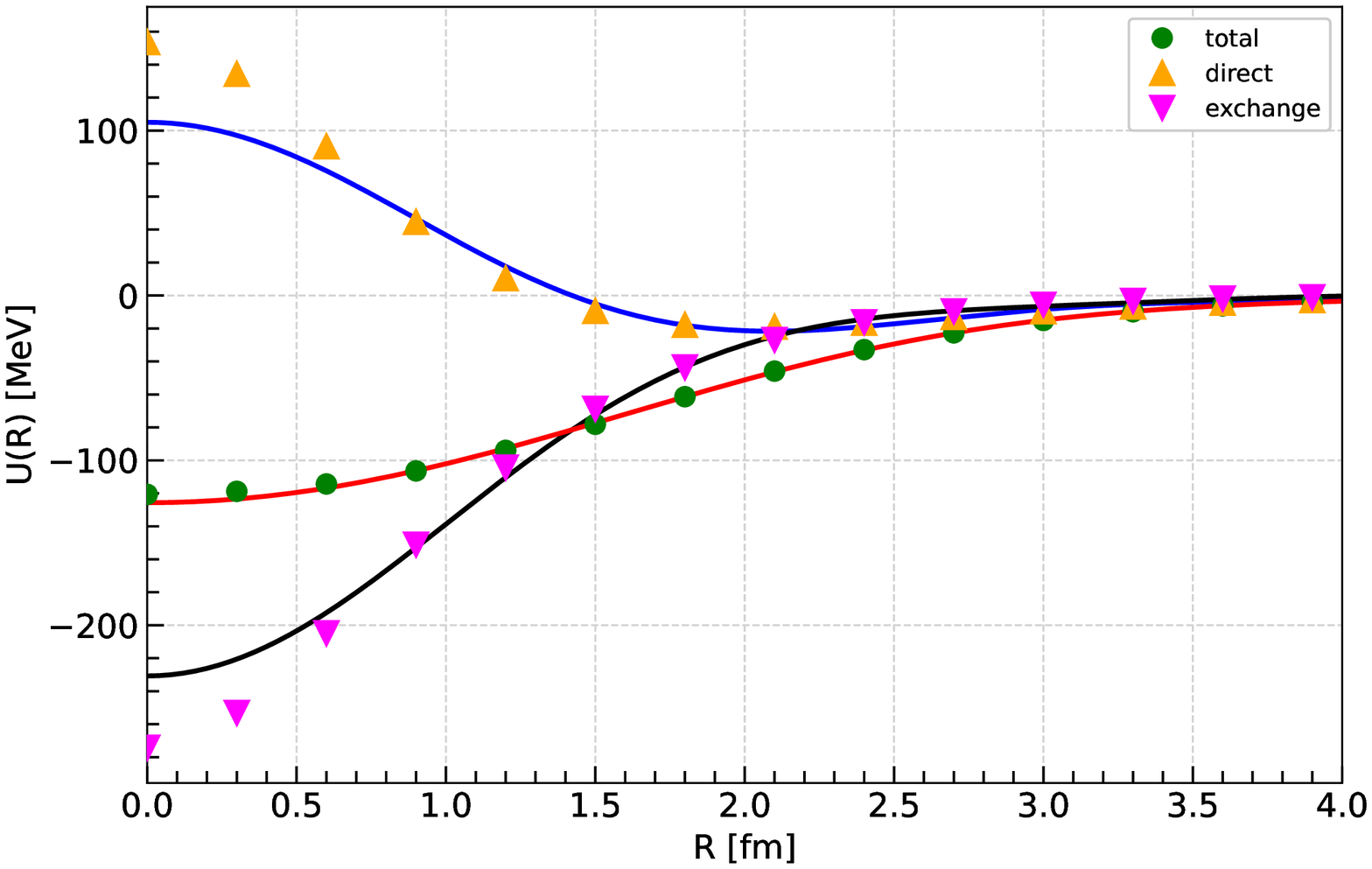}
		\caption{$q_{max}~=~3~\textrm{fm}^{-1}$}
		\label{fig:qb}
	\end{subfigure}
    \caption{The full circles, up triangles, and down triangles are the calculations of the total,
	direct and exchange parts of the analytically calculated double folding potentials, respectively.
	The solid lines are the computations using BiFold.}
	\label{fig:bf_vs_analytic}
\end{figure}

Both computations agree very well, as supported by the $\xi^2$ values in Table \ref{tab:mre} and shown 
in Fig. \ref{fig:qa}. 
The $\xi^2$ values are almost the same for both integration methods. There is a caveat to be careful of 
about this case. The effective NN interaction and the new nuclear matter distribution slowly go to 
zero when $r$ goes to infinity, contrary to the Gaussian-shaped density in Eq.~(\ref{eq:gauss}). 
This leads to the problem shown in Fig. \ref{fig:qb}, where $q_{max}~=~3~\textrm{fm}^{-1}$ is not enough 
to obtain a numerically accurate solution. 
Therefore it was necessary to raise $q_{max}$ to $10~\textrm{fm}^{-1}$ in this case while 
this value of $q_{max}$ is usually enough most of the time, as mentioned in Ref. \cite{COOK1982125}. 
This problem illustrates it is better to make a consistency check by raising the value of 
$q_{max}$ till the calculation reaches a saturation point where the potential does not change anymore.

\subsection{BiFold vs. DFPOT}
This test case compares the two codes for an $\alpha + ^{40}\textrm{Ca}$ elastic scattering system
where the energy of the $\alpha$ projectile is 141.7 MeV in the laboratory system. The effective NN interaction
is for both codes is the density-independent M3Y-Paris with the ZR exchange part given in Eqs.~(\ref{eq:parisD}) and (\ref{eq:parisZR}).
BiFold can perform the calculation for this case by changing  \postem{u\_m3y\_reid\_zr} to \postem{u\_m3y\_paris\_zr} 
in the twelfth line of Python code shown in Fig.  \ref{fig:bifold}.

\begin{figure}[!hbtp]
	\centering
	\includegraphics[width=0.6\textwidth]{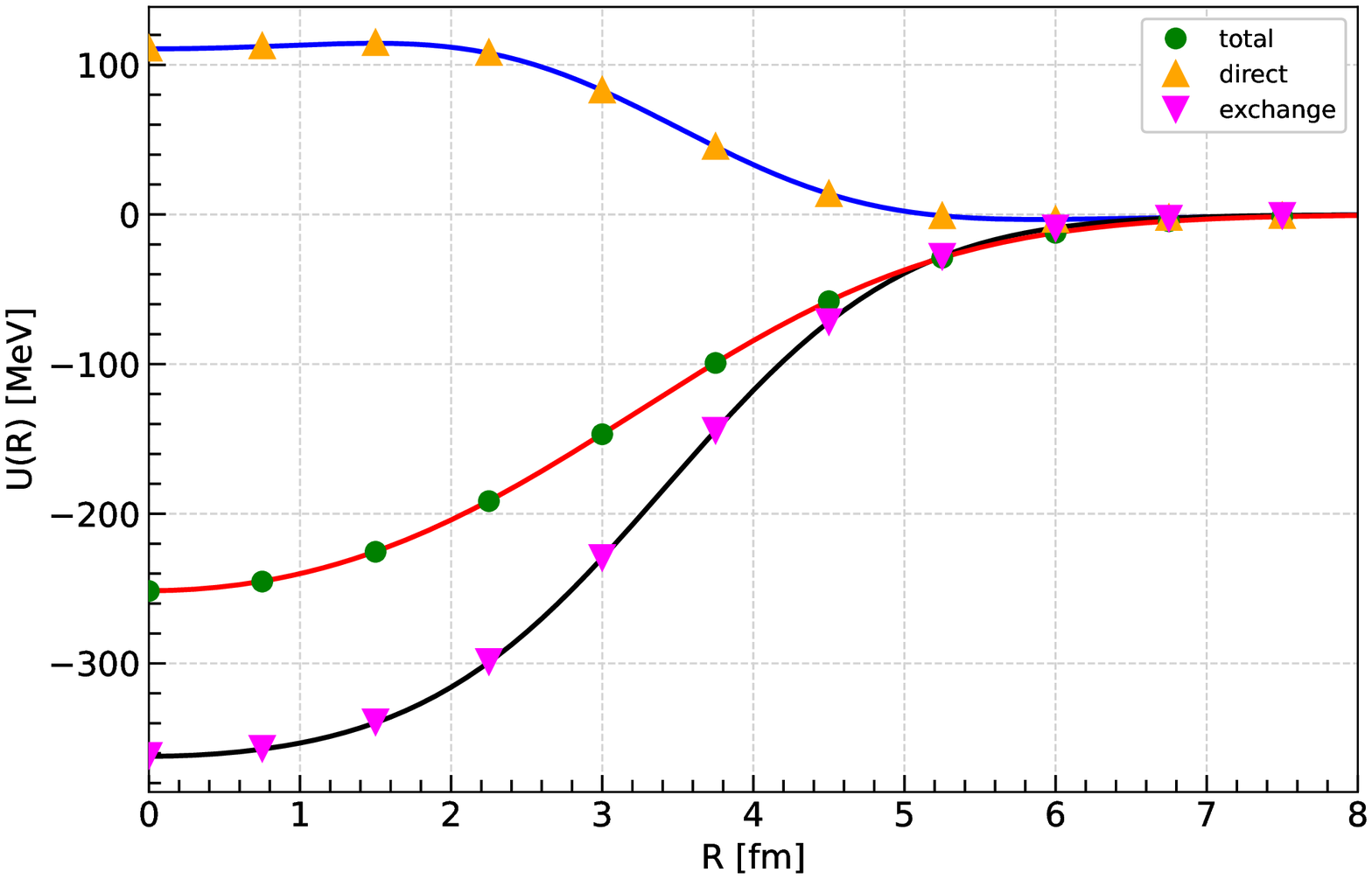}
	\caption{The full circles, up triangles, and down triangles are the calculations of the total,
		direct and exchange parts of the double folding potentials using DFPOT \cite{COOK1982125}, respectively. 
		The solid lines are the computations using BiFold.}
	\label{fig:bf_vs_dfpot}
\end{figure}

The nuclear matter density for $\alpha~-~\textrm{particle}$ is in Eq.~(\ref{eq:gauss}), and 
the density \cite{FARID1985525} of $^{40}\textrm{Ca}$ is
\begin{equation}
	\rho(r) = 0.169~\left[1 + \exp{\left(\frac{r-3.60}{0.523}\right)}\right]^{-1}~\textrm{fm}^{-3},
\end{equation}
with a msr value of $~\langle r^2\rangle = 11.553~\textrm{fm}^2$. As seen in Fig. \ref{fig:bf_vs_dfpot}, 
the computations of both codes are in very well agreement with each other, 
and the $\xi^2$ values in Table \ref{tab:mre} also support the claim.

\subsection{BiFold vs. Khoa \etal's calculation} 
The $^{16}\textrm{O}+^{16}\textrm{O}$ elastic scattering system at several incident energies was studied 
by Khoa \etal \cite{KHOA199574} using various density-dependent M3Y-Paris-based NN interactions.
The case chosen for the comparison is the one at 250 MeV incident energy. 
And the effective NN interaction used for the double folding potential is the BDM3Y1-Paris NN 
interaction with the FR exchange part.
The calculations of Khoa  \etal \cite{KHOA199574} were obtained by digitizing the related figure in 
the reference with the help of the programs Inkscape \cite{INKSCAPE} and Engauge Digitizer  \cite{MITCHELL20203941227} .

\begin{figure}[!hbtp]
	\centering
	\includegraphics[width=0.6\textwidth]{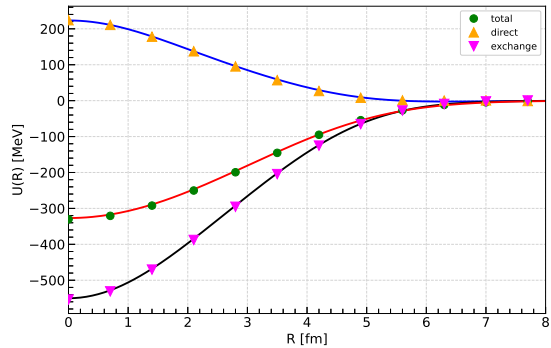}
	\caption{The full circles, up triangles, and down triangles are the calculations of the total,
		direct and exchange parts of the double folding potentials from Khoa \etal \cite{KHOA199574}, respectively. 
		The solid lines are the computations using BiFold.} 
	\label{fig:bf_vs_khoa}
\end{figure}

The nuclear matter density of the $^{16}\textrm{O}$ nuclei with the msr  value $~\langle r^2\rangle = 6.625~\textrm{fm}^2$ \cite{FARID1985525,KHOA199574} is
\begin{equation}
	\rho(r) = 0.181~\left[1 + \exp{\left(\frac{r-2.525}{0.45}\right)}\right]^{-1}~\textrm{fm}^{-3}.
\end{equation}
Since BDM3Y1 is a density-dependent NN interaction, the double folding integral in Eq.~(\ref{eq:df_ex})
has to be solved to obtain the potentials in Fig. \ref{fig:bf_vs_khoa}. 
This integral contains the Coulomb potential between the two nuclei through 
the local momentum $\vec K (\vec R)$ of the relative motion. 
The Coulomb potential modeled for uniformly charged spherical nuclei is
\begin{equation}
	U_{C}(R)=z_p z_t \frac{e^2}{4\pi\varepsilon_0} 
	\begin{cases}
		\frac{1}{R} & \left(R\geq R_{C}\right) \\ 
		\frac{1}{2 R_{c}}\left[3-\left(\frac{R}{R_{c}}\right)^{2}\right] & \left(R\leq R_{C}\right)
	\end{cases}
\end{equation}
used in the calculations where $z_p$, $z_t$ are proton numbers of the projectile and the target, respectively.
The Coulomb radius is $R_c = 1.405 \left(a_p^{1/3} + a_t^{1/3}\right)~\textrm{fm}$.

As can be seen from the $\xi^2$ values in Table \ref{tab:mre} and the potentials in Fig. \ref{fig:bf_vs_khoa}, 
BiFold is also consistent with the final reference work. 
It is important to note here that the comparison is made till 5.6~fm in this case since 
the resolution of the digitized figure \cite{KHOA199574} was not enough to recover the data with enough precision.





\section*{Declaration of competing interest}
The author declares that he has no known competing financial interests or 
personal relationships that could have appeared to influence the work reported in this paper.

\section*{Acknowledgements}

The author would like to thank \href{https://scholar.google.com/citations?user=Td0NnaIAAAAJ}{Prof. Dr. O. Bayrak} for stimulating discussions 
and useful comments on the manuscript.

\bibliographystyle{elsarticle-num}
\bibliography{paper}







\end{document}